\documentclass[aps, pra, reprint]{revtex4-1}
\usepackage{amsmath}
\usepackage{amssymb}
\usepackage{graphicx}
\usepackage{esint}
\usepackage{physics}
\usepackage{siunitx}
\usepackage[version=3]{mhchem}
\usepackage{placeins}
\usepackage{relsize}
\usepackage{array}
\usepackage{cleveref}
\Crefname{appsec}{Appendix}{Appendices}
\crefname{appsec}{appendix}{appendices}
\usepackage{comment}

\newenvironment{conditions}
  {\par\vspace{\abovedisplayskip}\noindent\begin{tabular}{>{$}l<{$} @{${}={}$} l}}
  {\end{tabular}\par\vspace{\belowdisplayskip}}

\newcommand{\acronym}[1]{\textsmaller{#1}}
\newcommand{\isreQsigma}{\ensuremath{Q_{[\sigma_{k}]}^{1}}}
\newcommand{\isreQtau}{\ensuremath{Q_{[\tau^{\text{ex}}]}^{1}}}
\newcommand{\isretauexfwd}{\ensuremath{\tau_{\text{fwd}}^{\text{ex}}}}
\newcommand\captiondescription[2]{\caption{#1 \newline #2}}

\begin{document}

\title{Spin Waves in Quantum Gases --- The Quality Factor of the Identical Spin Rotation Effect}
\author{L. Lehtonen}
\email{laanle@utu.fi}
\author{O. Vainio}
\author{J. Ahokas}
\author{J. J{\"a}rvinen}
\author{S. Sheludyakov}
\altaffiliation[Current affiliation:]{Institute for Quantum Science and Engineering}
\author{K.-A. Suominen}
\author{S. Vasiliev}
\affiliation {Department of Physics and Astronomy, University of Turku, 20014 Turku, Finland}

\author{V. Khmelenko}
\author{D. M. Lee}
\affiliation {Institute for Quantum Science and Engineering,  Department of Physics and Astronomy, Texas A\&M University, College Station, TX, 77843, USA}
\begin{abstract}
Our recent experimental work on electron spin waves in atomic hydrogen gas has 	prompted a revisit of the theory of the Identical Spin Rotation Effect (\acronym{ISRE}). A key characteristic determining the properties of the spin waves is the quality factor of \acronym{ISRE}. Unfortunately, calculating this quality factor takes some toil. In this paper we summarize some results of the \acronym{ISRE} theory in dilute gases. We also derive asymptotic formulae for the quality factor and examine their accuracy for hydrogen and \ce{^3 He}. 
  
\end{abstract}
\maketitle

\section{Introduction}

Studies of quantum gases have been an integral part of quantum optics for several decades. The ground state of spin-polarised atomic hydrogen forms an interesting four-level system that can be controlled by NMR (nuclear spin) and ESR (electron spin) methods. The role of atomic collisions, however, can be very different in this system compared to laser spectroscopy or cooling of atoms. The key point is that atomic collisions can lead to collective spin phenomena such as spin waves and their description with quasiparticles, the magnons. Thus it is both interesting and crucial to understand collisions in spin-polarised low-temperature hydrogen atoms. By making this study we wish to honour Prof. Dr. Wolfgang Schleich and his many significant contributions to quantum optics and beyond.

In the scattering of atoms, indistinguishablity and identicalness play a central role. Crucially, identical atoms in the same spin state experience interference effects, but identical atoms in orthogonal spin states behave as distinguishable atoms. The difference is particularly stark for fermions, where the difference between parallel and orthogonal spin states determines whether the lowest-order interaction is the partial p-wave or s-wave, respectively. 

The result is an effective spin-dependent interaction between the atoms, known as the Identical Spin Rotation Effect (\acronym{ISRE})\cite{LL1}. \acronym{ISRE} acts between identical atoms in neither parallel nor orthogonal spin states. Its effect is a rotation of the interacting spins around their sum, an inevitable consequence of the different phase shifts the different spin components of a superposition state experience. As an exchange effect, \acronym{ISRE} becomes more pronounced  as the wave function overlap becomes larger than the typical interaction range, being significant already in the quantum gas regime ($\Lambda_{\text{th}}> a_s$).

\acronym{ISRE} is intimately tied to transport phenomena such as heat conduction and spin diffusion. In particular, helical spin currents or spin waves have been predicted\cite{LL2, Bouchaud1985-Spinwaves} and observed for example in  nuclear spins\cite{Johnson1984, Bigelow1989} and in electron spins of atomic hydrogen \cite{Vainio2012, Vainio2015}. In electron spin spectra these spin waves modify the shape of the main absorption peak and create side peaks related to certain wave numbers. The temperature and density behavior of one of the peaks has also suggested that one may treat these spin waves as quasiparticles (magnons) which undergo Bose-Einstein Condensation \cite{Vainio2012}. Spin transport effects related with  
\acronym{ISRE} have also been observed in liquid \ce{^3 He}--\ce{^4 He} mixtures \cite{Gully1984}, \ce{^3 He} gas \cite{Nacher1984, Tastevin1985}, and in the cold gas of \ce{^{87} Rb} \cite{Fuchs2002, Deutsch2010, Maineult2012}. 

The main equations describing spin transport in quantum gases turned out to be identical with those for the degenerate Fermi liquids, e.g. mixtures of \ce{^3 He} in \ce{^4 He}, where the spin precession occurs due to an effective molecular field \cite{Leggett-Rice, Leggett_1970}. In fact, the theory of degenerate Fermi liquids was first used to predict and characterize spin waves in quantum gases \cite{Bashkin}. It was later shown\cite{Miyake1985} that this similarity of the \acronym{ISRE} theory and the Leggett-Rice theory in Fermi liquids is not a coincidence, but a consequence of the same physical origin of spin transport phenomena in these systems. 

The quality factor of the spin waves is central in characterizing the region where spin waves persist. It is a measure of the persistence of the spin waves against homogenizing diffusion, and can be given as a ratio of the spin wave frequency to their time decay constant. Generally it is related to the spin-wave quality factor $\mu$, a ratio measuring the effect of \acronym{ISRE} to classical diffusion. In the case of the spin-1/2 gas the quality factor of spin waves is given by $|\mu S|$, where $S$ stands for the spin polarization density of the unperturbed spin gas.  For higher spins one may have different results \cite{spin-pol-deut}. Couplings to other degrees of freedom may also significantly reduce the  actual quality factor \cite{Bouchaud1985-Spinwaves}. These issues are elaborated on in \cref{isre-types}.

The calculation of the spin-wave quality factor $\mu$ is quite complicated\cite{LL3}, generally involving various scattering quantities at different momenta and their averages. However at low temperatures only the s-wave scattering contributes significantly to the interactions (for both bosons and fermions, as \acronym{ISRE} ultimately occurs between non-parallel states). From heuristic arguments it has been known that $\mu \sim \frac{\Lambda_{\text{th}}}{a_s}$ as $T \to 0$, where the right-hand side is simply proportional to the 'quantumness' of the gas. The asymptotic limit of $\mu$ was first derived in \cite{Miyake1985}; in  \cref{sec:asymptotic-expression-derivation}, we repeat this derivation with more detail. For comparison, an expression with higher-order terms is also derived. With this as a basis, we derive the asymptotic limit for $\mu$ of electron spin waves based on Bouchaud and Lhuillier\cite{Bouchaud1985-Spinwaves}, who considered
specifically the case of $b$-$c$--coherence in atomic hydrogen and treat hydrogen explicitly as composed of a nucleus and an electron.
Finally, in \cref{sec:mu-for-different-species} the accuracy of these expressions in compared against a numerical calculation of $\mu$ for hydrogen and \ce{^3 He}.

\section{Types of ISR Equations}\label{isre-types}
The \acronym{ISR} equation is a spin diffusion equation which accounts for spin currents arising from symmetrization of wave functions. The first step in its derivation is to derive the relevant scattering cross section(s), first done by Pinard and Lalo\"{e}\cite{pinard-pauli} and later repeated in \cite{LL1, Bouchaud1985-Boltzmann}.  In the second step, the cross section is used in a Boltzmann equation which is subsequently solved using a Chapman-Enskog expansion, which looks at a small perturbation around the equilibrium spin polarization density $\vec{S}$. This gives an expression for the spin current \cite{LL2, Bouchaud1985-Spinwaves}, which in conjunction with the equation for precessing spin $\pdv{\vec{S}}{t}+\nabla\cdot\vec{J}=\gamma\vec{S}\times\vec{B}$ leads to the \acronym{ISR} equation for the transverse spin (polarization) density $S_x+i S_y$: 
\begin{equation*}
\pdv{S_{+}}{t}+i\gamma H_{z}S_{+}=D_{0}\frac{1-i\mu\epsilon S_{z}}{1+\mu^{2}S^{2}}\nabla^{2}S_{+},
\end{equation*}
where
\begin{conditions}
 D_0     &  spin diffusion coefficient \\
 n     &  gas number density \\   
 \varepsilon &  $\pm 1$ for bosons/fermions \\
 \mu     &  spin wave quality factor \\   
 S     &  magnitude of the longitudinal ($z$) spin polarization \\
 S_+ & transverse spin polarization \\
 H_z & component of $\vec{H}$ parallel to $S$ \\
 \gamma & gyromagnetic ratio of the electron or the nucleus.
\end{conditions}
For a highly-polarized gas $S \approx S_z$, and the equation simplifies slightly to 
\begin{equation}
\pdv{S_{+}}{t}+i \gamma B_{z}S_{+}=D_{0}\frac{1}{1+i\epsilon \mu S}\nabla^{2}S_{+} \label{LL-ISRE}.
\end{equation}
This is the \acronym{ISR} equation based on Lhuillier and Lalo\"{e} (LL)\cite{LL2}, and it is valid for spin-½ systems with negligible couplings to other degrees of freedom, such as nuclear spins of atoms with 'frozen' electron spins (i.e. the electron spins are fully polarized and the electrons are bound to their respective atoms during collisions). 

Lhuillier and Lalo\"{e} used spin polarization density in their derivation instead of magnetization; further, they chose their axes so that the $S_z$($\approx S$) is always parallel to the positive $z$-axis. Specifically this means that $0 \leq S_z \leq 1$.  A less obvious consequence is that the sign of $\gamma$ plays no role in the equation. In particular for electrons $\gamma < 0$, however the spins are aligned against the field.  As $S_z$ by definition always points in the positive $z$-direction, in order for them to be aligned against the magnetic field, one must flip the direction of $H_z$; this exactly cancels the sign of $\gamma$ (see also \cref{gamma-sign}). Lastly, if one were to flip the polarization/magnetization of the gas, the correct way to account for it in the \acronym{ISRE} equation would be to flip the magnetic field. 

Bouchaud and Lhuillier\cite{Bouchaud1985-Spinwaves} considered atomic hydrogen taking into account both nuclei and electrons  (in fact some of their results are general to atoms with one valence electron\cite{Bouchaud1985-Boltzmann}) and allow electrons to jump from nucleus to the other during the collision. Specifically they consider the cases of the $0$-$0$ coherence in weak magnetic field and the $b$-$c$ coherence in strong magnetic field. For both cases they arrive at essentially the same \acronym{ISR} equation but in the former case with a drastically modified quality factor $Q$, which turns out to significantly limit the observability of $0$-$0$ spin waves: 
\begin{equation}
Q=\frac{\mu S_{z}}{1+\frac{1+\mu^{2}S_{z}^{2}}{D_{z}k^{2}}T_{2}^{-1}}
\end{equation}
For the $b$-$c$ coherence they obtain
\begin{equation}
\pdv{z}{t}=\frac{D_{z}^{*}\left(I\right)}{1+i\left(\mu_{1}^{*}M+\mu_{2}^{*}S\right)}\nabla^{2} z \label{bouchaud-isre-eq}, 
\end{equation}
where $M, S$, and $I$ are various quantities characterizing polarization (see \cref{sec:bouchaud-mu}). This equation has no $\varepsilon$s because they're already included in the calculation of the $\mu$s and the relationship is not as simple as it is for Lhuillier and Lalo\"{e}.

Earle\cite{spin-pol-deut} considered the spin-1 case of deuterium nuclei and following Lhuillier and Lalo\"{e} derived the \acronym{ISR} equations for various spin waves, but with $\mu S$ replaced by $2 \mu A_{\pm2}$ for $\gamma \leftrightarrow \alpha$ transition and $\epsilon \mu (S_0- \sqrt{6} A_0)$ for $\gamma \leftrightarrow \beta$ spin waves ($A_i$s are the components of the nuclear quadrupolar alignment tensor). 

Generally the \acronym{ISR} equation has the form
\begin{equation}
  \pdv{S_{+}}{t}=\frac{D_{0}}{1+ i\mu_{\text{eff}}S}\nabla^{2}S_{+}-i\gamma H_{z}S_{+}
  \label{general-isre-eq}
\end{equation}
with various expressions substituting for $\mu_{\text{eff}}$ and depending on approach, $S$ may be
positive (LL) or assume even negative value (Bouchaud and Lhuillier). 

\subsection{Trapping Magnons}
The \acronym{ISR} equation is mathematically similar to Schr\"{o}dinger equation
with magnetic field in the role of potential. While physically the \acronym{ISR}
equation is not in any obvious way connected to the energy of spin waves, nonetheless
one may use the intuition from Schr\"{o}dinger equation to say something about the
spin waves. So as one can speak of trapping quantum systems in a potential,
one may speak of spin waves being trapped by the potential (that is, the magnetic field).
Whether the spin waves are attracted to potential minima or maxima depends essentially only
on the sign of $\mu$. For example, consider $|\mu|\gg 1$: one may then write the
\acronym{ISR} equation as
\begin{equation}
-i\dv{S_+}{t}=-\underbrace{\frac{D_{0}}{\mu}}_{\frac{\hbar^{2}}{2m}}\nabla^{2}S_+ \underbrace{-\left|\gamma H_{z} \right|}_{V\left(r\right)}S_+.
\end{equation}
(Here the modulus of $\gamma H_z$ is used to emphasize the fact that the sign of
$\gamma$ is not relevant.) This differs from a Schr\"{o}dinger equation only by the sign
of $\dv{S_+}{t}$; taking complex conjugate would recover a Schr\"{o}dinger equation for
the conjugate $S_{+}^{*}$, a particle with mass $m=\frac{\hbar^{2}\mu}{2D_{0}}$.
What matters is that the sign of the `kinetic' term is the same as in Schr\"{o}dinger equation.
Now for $\mu>0$, a strong magnetic field corresponds to a potential minimum,
so the resulting spin waves should concentrate in regions of strong magnetic field.
Spin waves in stronger magnetic field would also have higher precession frequency:
as the mode number increases, the frequency should decrease as the modes move out
of the potential to regions of weaker magnetic field. An increasing frequency
spectrum would be observed for $\mu < 0$ or flipped magnetization
($-\left|\gamma H_{z} \right|\to \left|\gamma H_{z} \right|$), but not for a
change in the sign of $\gamma$ as previously explained.

 \section{The Asymptotic Expressions} \label{sec:asymptotic-expression-derivation}
 The \acronym{ISRE} parameter $\mu$ characterizes the ratio of transverse rotation of \acronym{ISRE} to normal spin diffusion tending to homogenize the gas. It depends on three different cross sections  (momentum $k=\norm{\vec{k}}$): the usual scattering cross section $\sigma_k(\theta)$ of scattered atoms, an interference term $\tau^{\text{ex}}_{\text{fwd}}(k)$ for transmitted atoms, and an interference term for scattered particles $\tau^{\text{ex}}_k (\theta)$ \cite{LL1}. From these one may obtain the angle-integrated cross sections
\begin{equation*}
Q^t_{[\sigma]}(k) = 2 \pi \int_0^{\pi} \sin \theta (1-\cos^t \theta) \sigma (\theta)
\end{equation*}
The phase-shift expansion of the $T$-matrix gives the expressions of \cite{LL3}:
\begin{align}
Q_{[\sigma_{k}]}^{1}= & \frac{4\pi}{k^{2}}\sum_{l=0}^{\infty}\left(l+1\right)\left(\sin\left(\delta_{l}-\delta_{l+1}\right)\right)^{2}\nonumber \\
Q_{[\tau^{\text{ex}}]}^{1}= & \frac{8\pi}{k^{2}}\sum_{l=0}^{\infty}\left(-1\right)^{l}\left(l+1\right)\sin\left(\delta_{l}-\delta_{l+1}\right)\sin\left(\delta_{l}\right)\sin\left(\delta_{l+1}\right)\nonumber \\
\tau_{\text{fwd}}^{\text{ex}}= & \frac{2\pi}{k^{2}}\sum_{l=0}^{\infty}\left(-1\right)^{l}\left(2l+1\right)\sin\left(2\delta_{l}\right)\nonumber 
\end{align}
Here $\delta_l$ is the $l$-wave phase shift. From these, using the collision integrals 
\begin{align}
\Omega_{[\alpha]}^{(t,s)}= & \frac{1}{\sqrt{\pi m\beta}}\int_{0}^{\infty}e^{-\gamma^{2}}\gamma^{2s+3}Q_{[\alpha]}^{t}\left(k=\sqrt{\frac{m}{\beta}}\frac{\gamma}{\hbar}\right)\mathrm{d}\gamma\label{eq:omega-alpha}\\
\Xi_{[\tau_{\text{fwd}}^{\text{ex}}]}^{(s)}= & \frac{1}{\sqrt{\pi m\beta}}\int_{0}^{\infty}e^{-\gamma^{2}}\gamma^{2s+3}\tau_{\text{fwd}}^{\text{ex}}\left(k=\sqrt{\frac{m}{\beta}}\frac{\gamma}{\hbar}\right)\mathrm{d}\gamma\label{eq:chi-tau},
\end{align}
one arrives to the definition of $\mu$: 
\begin{equation*}
\mu=  \frac{\Omega_{[\tau^{\text{ex}}]}^{(1,1)}+\Xi_{[\tau_{\text{fwd}}^{\text{ex}}]}^{(1)}}{\Omega_{[\sigma_{k}]}^{(1,1)}}\nonumber 
\end{equation*}
Assuming that at low momenta the phase shift behaves asymptotically as $\delta_{l}  =-\left(ka_{l}\right)^{2l+1}+n\pi$ justifies the definition of $l$-wave scattering length:
\begin{equation*}
a_{l} = -\left(\lim_{k \to 0}\frac{\tan\left(\delta_{l}\left(k\right)\right)}{k^{2l+1}}\right)^{\frac{1}{2l+1}}
\end{equation*}
For $l=0,1$ this usually works,  but for higher partial waves the scattering length defined thus may not be finite. However, the phase shift may behave as $\sim (k a_l )^q $ for some $q$, in which case higher scattering lengths may be defined by suitably adjusting the definition.

Expanding the angle-averaged quantities to second order in $k$ gives,
\begin{align}
Q_{[\sigma_{k}]}^{1}= & 4\pi a_{0}^{2}-4\pi a_{0}k^{2}\left(\frac{a_{0}^{3}}{3}+2a_{1}^{3}\right)+\mathcal{O}\left(k^{4}\right) \label{eq:q1-sigma-appr}\\
Q_{[\tau^{\text{ex}}]}^{1}= & \mathcal{O}\left(k^{3}\right)\nonumber \\
\tau_{\text{fwd}}^{\text{ex}}= & -\frac{4\pi a_{0}}{k}+4\pi k\left(\frac{2}{3}a_{0}^{3}+3a_{1}^{3}\right)+\mathcal{O}\left(k^{3}\right) \label{eq:tau-ex-appr}
\end{align}

One could also take terms up to first order in $k$, but most likely owing to the rational form of $\mu$ these seem to be less accurate. Higher order
terms may depend on $a_{2}$ which, as mentioned before, generally isn't finite, and one may have to consider the sign of $Q_{[\tau_{\text{fwd}}^{\text{ex}}]}^{1}$
due to $n\pi$ term of $\delta_{l}$, which turns out not to be an
issue for $Q_{[\sigma_{k}]}^{1}$ (square of sine) and $\tau_{\text{fwd}}^{\text{ex}}$
(always a multiple of $2\pi$). 

In the next step the above expressions are integrated with a Gaussian
over all $k$. The validity of this approximation is discussed in
\cref{validity-regimes}. The results are

\begin{align*}
\Omega_{[\sigma_{k}]}^{(1,1)}= & 4\pi a_{0}^{2}-\frac{4\pi a_{0}^{4}m}{\beta\hbar^{2}}-\frac{24\pi a_{0}m}{\beta\hbar^{2}}a_{1}^{3}\\
\Xi_{[\tau_{\text{fwd}}^{\text{ex}}]}^{(1)}= & -\frac{3\pi^{\frac{3}{2}}a_{0}\hbar}{2\sqrt{m}}\sqrt{\beta}+\frac{5\pi^{\frac{3}{2}}a_{0}^{3}\sqrt{m}}{2\sqrt{\beta}\hbar}+\frac{45\pi^{\frac{3}{2}}a_{1}^{3}\sqrt{m}}{4\sqrt{\beta}\hbar}.
\end{align*}

Taking the first term of both expressions (with $\Lambda=\sqrt{\frac{h^{2}\beta}{2\pi m}}$), the asymptotic
behavior is
\begin{equation} \label{eq:first-order-mu}
  \mu=  -\frac{3\sqrt{2}}{16}\left(\frac{\Lambda}{a_{0}}\right).
\end{equation}
The same result was obtained earlier in \cite{Miyake1985}. As can be seen, the sign of $\mu$ at low temperatures is determined
by the $s$-wave scattering length.

Using all the terms of the collision integrals above,  one arrives to the first-order $\mu$:
\begin{multline} \label{eq:second-order-mu}
\mu=-\frac{-3\sqrt{2}\Lambda^{3}}{16\Lambda^{2}a_{0}-32\pi a_{0}^{3}-192\pi a_{1}^{3}}+\frac{5\sqrt{2}\pi\Lambda a_{0}^{2}}{8\Lambda^{2}a_{0}-16\pi a_{0}^{3}-96\pi a_{1}^{3}}\\+\frac{45\sqrt{2}\pi\Lambda a_{1}^{3}}{16a_{0}\left(\Lambda^{2}a_{0}-2\pi a_{0}^{3}-12\pi a_{1}^{3}\right)}. 
\end{multline}


\subsection{Asymptotic Behaviour of $\mu$ in Bouchaud and Lhuillier's Treatment of $b$-$c$ Coherence in Atomic Hydrogen}\label{sec:bouchaud-mu}
Bouchaud and Lhuillier's treatment of the \acronym{ISRE} problem is far more detailed compared to Lhuillier and Lalo\"{e}'s, and as a result the calculations are even more cumbersome. The expressions for the $\mu^*$ factors in \eqref{bouchaud-isre-eq} are given by the following formulae: 

\begin{align} \label{bouchaud-formulae}
  D_{0}= &	\left[\frac{8nm}{3kT}\right]\left[\Omega_{\left[\sigma_{d}\right]}^{\left(1,1\right)}-\left(\frac{1+I}{2}\right)\Omega_{\left[\sigma_{dt}\right]}^{\left(1,1\right)}\right. \\
  +& \left(1-I\right)\Omega_{\left[\sigma_{dt}^{ex}\right]}^{\left(1,1\right)}-\tilde{\Omega}_{\left[\sigma_{dt}^{ex}\right]}^{\left(1,0\right)}+
  \frac{1}{2}\tilde{\Omega}_{\left[\sigma_{t}\right]}^{\left(0,0\right)}\\
+&\left.\left(1-I\right)\tilde{\Omega}_{\left[\sigma_{t}^{ex}\right]}^{\left(1,0\right)}+\frac{3I}{2}\Omega_{\left[\sigma_{t}^{ex}\right]}^{\left(1,0\right)}+\left[\frac{1-I}{2}\right]\Omega_{\left[\sigma_{d}^{ex}\right]}^{\left(1,1\right)}\right] \\
\frac{\mu^*_{1}}{D_{0}}=&	\left[\frac{8nm}{6kT}\right]\left[\Xi_{\tau_{d}^{bwd}}^{1}-\tilde{\Omega}_{\left[\tau_{d}^{ex}\right]}^{\left(1,1\right)}-\tilde{\Omega}_{\left[\tau_{t}^{ex}\right]}^{\left(1,1\right)}\right.\\
+&\left. 3\Xi_{\tau_{t}^{fwd}}^{0}-\Xi_{\tau_{t}^{fwd}}^{1}+6\tilde{\Omega}_{\left[\tau_{dt}\right]}^{\left(0,0\right)}-2\tilde{\Omega}_{\left[\tau_{dt}\right]}^{\left(1,1\right)}\right]\\
  \frac{\mu^*_{2}}{D_{0}}=&	-\left[\frac{8nm}{6kT}\right]\left[\Xi_{\tau_{t}^{bwd}}^{1}+3\Xi_{\tau_{t}^{fwd}}^{0}\right. \\
  +& \left. 2\tilde{\Omega}_{\left[\tau_{dt}\right]}^{\left(1,1\right)}+6\tilde{\Omega}_{\left[\tau_{dt}\right]}^{\left(0,0\right)}\right].
\end{align}
The quantities in the formulae are similar to those used in the previous section. A more comprehensive summary is given in \cref{crosssections}.

Using the SymPy Python package\cite{sympy} to perform the expansions to first order, the expressions for the $\mu$s one obtains are
\begin{align}
\mu_{1}^{*}=&	-\frac{3\sqrt{2}\Lambda a_{g}}{4Ia_{g}^{2}-16Ia_{g}a_{u}-4Ia_{u}^{2}+10a_{g}^{2}+4a_{g}a_{u}+34a_{u}^{2}} \\
\mu_{2}^{*}=&	\frac{9\sqrt{2}\Lambda\left(a_{g}-a_{u}\right)}{8Ia_{g}^{2}-32Ia_{g}a_{u}-8Ia_{u}^{2}+20a_{g}^{2}+8a_{g}a_{u}+68a_{u}^{2}} \\
\mu=&	M\mu_{1}^{*}+S\mu_{2}^{*} \label{eq:bouchaud-asymptotic}
\end{align}

with 
\begin{align*}
I=&	\frac{n_{a}+n_{d}-n_{b}-n_{c}}{n}\quad\left(\text{nuclear polarization}\right) \\
S=&	\frac{n_{d}+n_{c}-n_{a}-n_{b}}{n}\quad\left(\text{electron polarization}\right) \\
M=&	\frac{n_{c}-n_{b}}{n}.
\end{align*}
$a_{g}$ and $a_{u}$ correspond to singlet and triplet potential scattering lengths, and the $n$s are the number densities of different spin states of atomic hydrogen in strong fields. For a gas consisting of pure $b$-state, $I=S=M=-1$ and one obtains
\begin{align*}
  \mu = & -\mu^*_1-\mu^*_2 = -\left( - \frac{3 \sqrt{2} \Lambda a_{g, 0}}{6 a_{g, 0}^{2} + 20 a_{g, 0} a_{u, 0} + 38 a_{u, 0}^{2}} \right. \\
  + & \left.\frac{9 \sqrt{2} \Lambda \left(a_{g, 0} - a_{u, 0}\right)}{12 a_{g, 0}^{2} + 40 a_{g, 0} a_{u, 0} + 76 a_{u, 0}^{2}}\right).
\end{align*}
To compare with Lhuillier and Lalo{\"e} we artificially set $a_{g, 0}=0$: one is left with
$-\left(- \frac{9 \sqrt{2} \Lambda}{76 a_{u, 0}}\right)=\approx 0.12 \frac{\sqrt{2} \Lambda}{a_{u, 0}}$,
to be compared with LL's $\varepsilon \mu S \approx -0.19 \frac{\sqrt{2} \Lambda}{a_ 0}$.

\section{Comparison of Exact and Asymptotic Curves for $\mu$} \label{sec:mu-for-different-species}

\begin{figure}
\begin{centering}
\includegraphics[width=0.95\columnwidth]{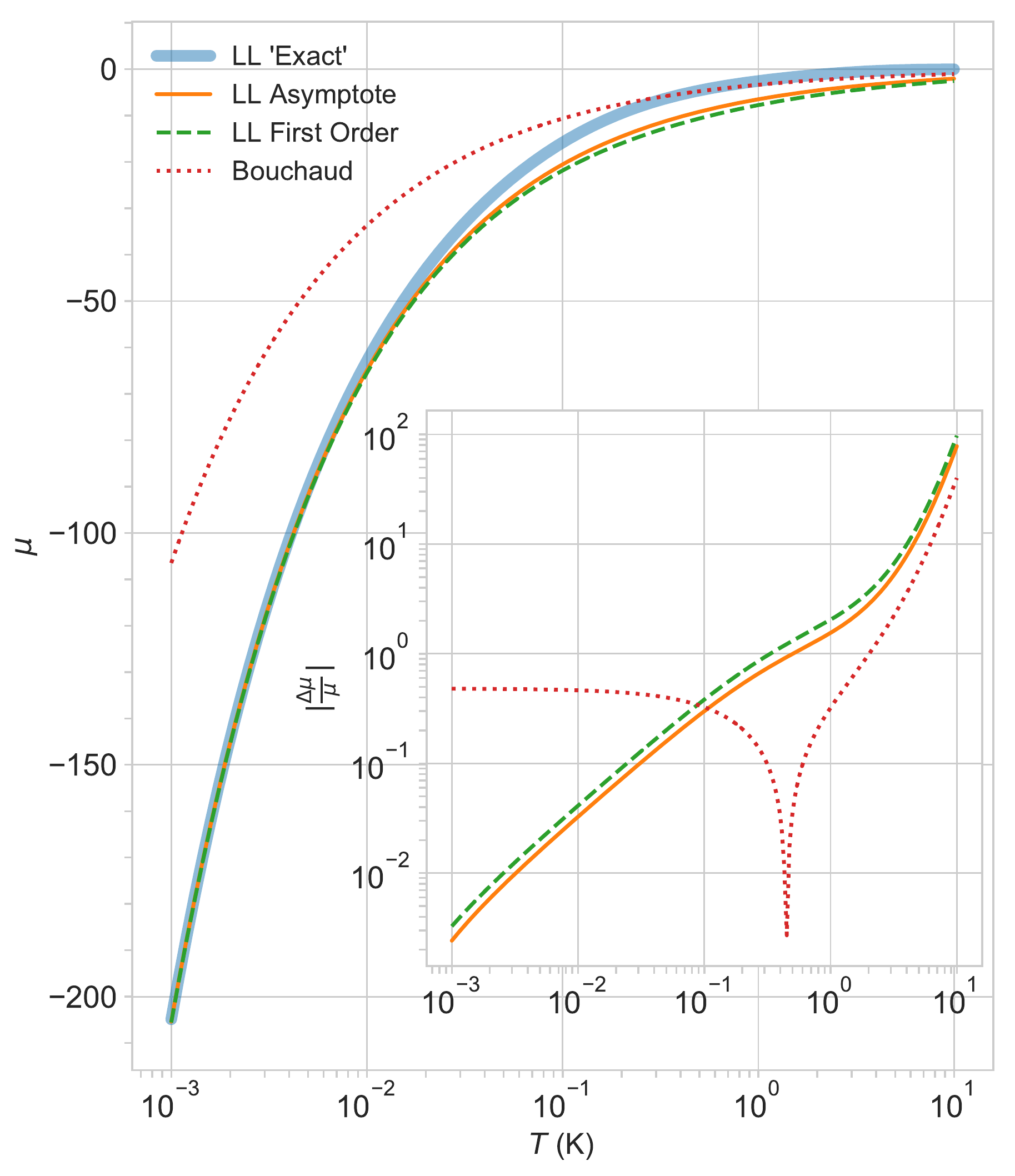}
\end{centering}
\captiondescription{Accuracy of the Asymptotic Approximations for Hydrogen}
{The figure shows the `exact' calculation, the asymptotic formula~\cref{eq:first-order-mu}, the first-order formula~\cref{eq:second-order-mu}, and the asymptotic Bouchaud formula~\cref{eq:bouchaud-asymptotic} for~$\mu$ in hydrogen.
The inset shows the relative error from the 'exact' value.  The discrepancy is around 30\% for $T\sim0.1\text{K}$ for the LL formulae.
The Bouchaud and Lhuillier formula ($\mu=\mu_1^*+\mu_2^*$) differs more at low temperatures.}\label{hydrogen-mu-curves}

\end{figure}

\begin{figure}
\begin{centering}
\includegraphics[width=0.95\columnwidth]{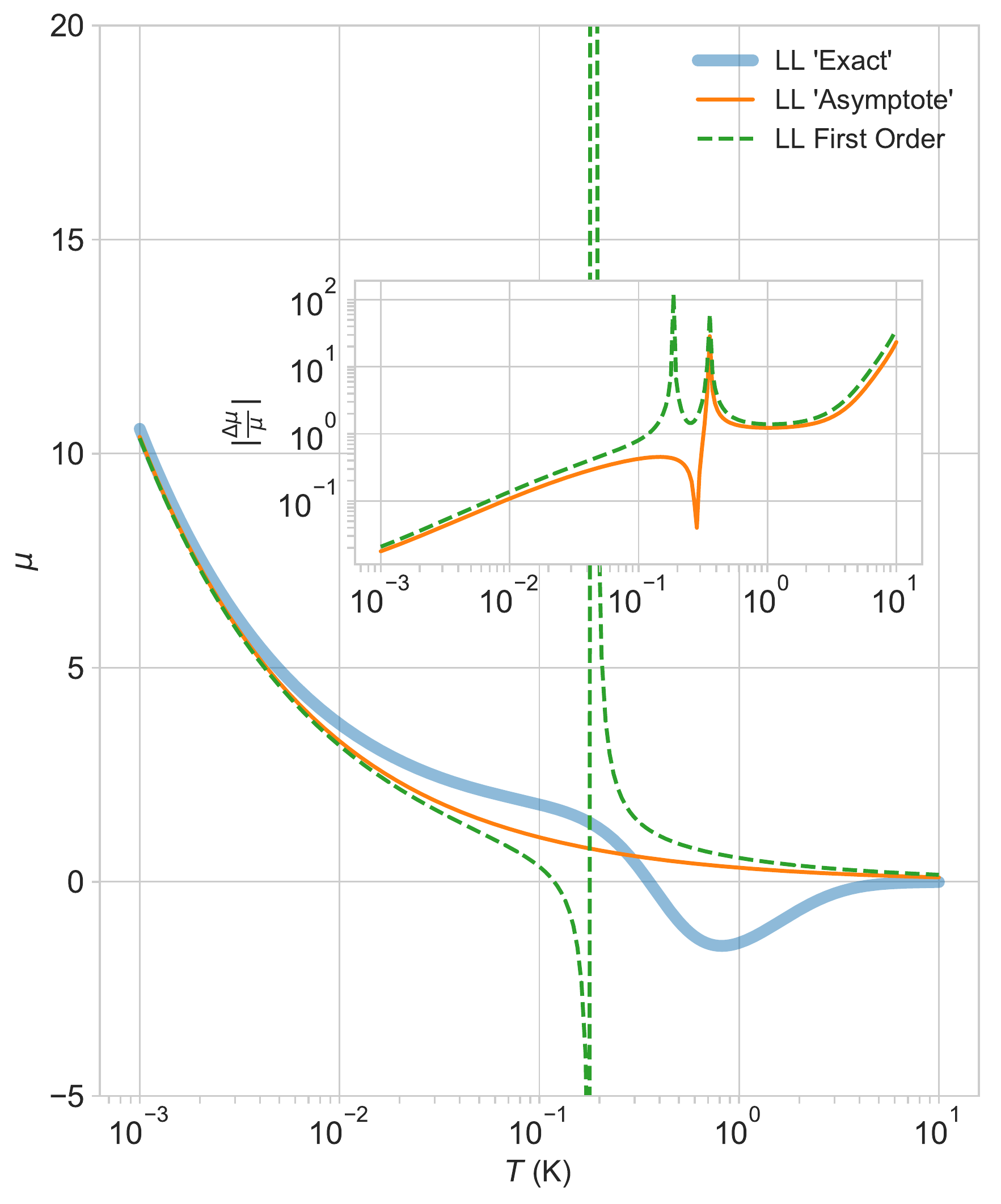}
\end{centering}
\captiondescription{Accuracy of the Asymptotic Approximations for \ce{^3 He}}{
The figure shows the `exact' calculation, the asymptotic formula~\cref{eq:first-order-mu},
and the first-order formula~\cref{eq:second-order-mu} ~$\mu$ in \ce{^3 He}.
The inset shows the relative error from the 'exact' value.  The discrepancy is around 10\% for $T\sim0.01\text{K}$. $\mu$ changes sign
which leads to the divergence of the first-order result.}\label{helium-mu-curves}

\end{figure}

\Cref{hydrogen-mu-curves} and \cref{helium-mu-curves} show $\mu$ calculated for
hydrogen's triplet potential ($\ce{b ^3 \Sigma^+_u}$) and \ce{^3 He}. 
The figures also show a comparison between the asymptotic formulae and more comprehensive 
calculations in the fashion of \cite{LL3} (`exact'). A trend that seems to emerge
from these examples is that the first-order expression differs more from the exact
result than the zeroth-order asymptotic formula. Given the poorness of the results especially for \ce{^3 He},
it seems likely that the first-order formula is a poor approximation of $\mu$, though it remains possible
it is the `exact' $\mu$ which is inaccurate.

For hydrogen, a refined Kolos-Wolniewicz potential\cite{Jamieson2000} was used to calculate
$a^u_{0}=\SI{0.71}{\angstrom}$ and $a^u_{1}=\SI{-2.70}{\angstrom}$ with the
variable phase method\cite{Calogero1968}; the results are in good agreement with other
calculations\cite{Joudeh2013}. For the singlet scattering length in the Bouchaud and Lhuillier approximation,
$a^g_0=\SI{0.16}{\angstrom}$ was used\cite{Jamieson2010}.
The required phase shift curves for the `exact' $\mu$ were calculated using a combination of the variable phase method
and a version of the usual solution-matching method\cite{Wei2006}.
The resulting asymptotic and first-order curves are seen to follow the `exact' curve, although
the relative discrepancy even at \SI{0.1}{\kelvin} is around 30\% for both. Further, the first-order
expression is clearly less accurate. The Bouchaud and Lhuillier formula $\mu=\mu_1^*+\mu_2^*$
fares better at higher temperatures (above \SI{0.1}{\kelvin}) compared to either LL results.
However the situation changes at about \SI{0.1}{\kelvin} where the asymptotic Bouchaud and Lhuiller
curve departs from the other curves. On one hand this speaks of the applicability
of the LL treatment in many contexts, on the other hand it shows that Bouchaud's more detailed approach differs
from the more general treatment of LL to a degree which cannot be explained by the LL theory. 
In particular it would seem to suggest differences between electron
spin waves ($b$-$c$--coherence) and nuclear spin waves ($a$-$d$--coherence) in hydrogen.
With Bouchaud and Lhuillier's definition of $M<0$ for
hydrogen gas in pure $b$-state and LL's definition $M>0$ always, the $\mu_{\text{eff}} M$ should have different sign
for these two approaches.

 For \ce{^3He}, $a_{0}=\SI{-8.0592}{\angstrom}$ and $a_{1}=\SI{-3.024}{\angstrom}$, calculated
 from a Lennard-Jones potential\cite{LL3}. Once again both curves approximate
 the `exact' result equally well until the discontinuity; there the asymptotic
 formula fares better though neither correctly approximates the behavior.


\section{Conclusions}
We derived asymptotic expressions for the quality factor of the \acronym{ISRE} nuclear spin waves in quantum gases of atomic hydrogen and \ce{^3 He} using the theory of Lhuillier and Lalo{\"e} based on scattering of identical particles with spin-$\frac{1}{2}$. For electron spin waves in atomic hydrogen we used a more accurate treatment of Bouchaud and Lhuillier where the true four-particle nature of the scattering (two electrons, two nuclei) is considered. The  quality factor parameter $\mu$ was calculated with first and second order approximations. Comparing the asymptotic values of $\mu$ with results of exact numerical calculations we found that they agree well with each other within the experimentally accessible range of temperatures \SIrange{0.1}{1}{\kelvin}, and diverge at low temperatures as $\Lambda/ a_0$.

\appendix
\crefalias{section}{appsec}
\section{The Effect of the Sign of $\gamma$}\label{gamma-sign}
Assuming the time-dependence of \acronym{ISRE} spin waves to be given by $e^{i \Omega t}$ (with $\Omega$ generally being a complex number), \cref{general-isre-eq} becomes
\begin{equation}\label{general-isre-omega}
  i \Omega S_{+}=\frac{D_{0}}{1\pm i\mu_{\text{eff}}}\nabla^{2}S_{+}-i\gamma H_{z}S_{+}.
\end{equation}
To make explicit the dependence of $S$ on $\gamma$ we substitute $S=\frac{2}{\gamma \hbar} M$; the substitution cancels out everywhere but the denominator on the right.

\begin{equation}
  i \Omega M_{+}=\frac{D_{0} \gamma \hbar}{\hbar \gamma \pm 2 i\mu_{\text{eff}} M}\nabla^{2}M_{+}-i\gamma H_{z}M_{+}.
\end{equation}

Substituting $\gamma \to -\gamma$ leads to

\begin{align}
  i \Omega M_{+}& =\frac{- D_{0} \gamma \hbar}{- \hbar \gamma \pm 2 i\mu_{\text{eff}} M}\nabla^{2}M_{+}+i\gamma H_{z}M_{+} \\
  & = \frac{ D_{0} \gamma \hbar}{ \hbar \gamma \mp 2 i\mu_{\text{eff}} M}\nabla^{2}M_{+}+i\gamma H_{z}M_{+}.
\end{align}
Then we take the  complex conjugate:
\begin{equation}
  i \overbrace{(-\Omega^*)}^{\widetilde{\Omega}} M_{+}^*=\frac{D_{0} \gamma \hbar}{\hbar \gamma \pm 2 i\mu_{\text{eff}} M}\nabla^{2}M_{+}^*-i\gamma H_{z}M_{+}^*.
  \label{flipped-conjugate-isre}
\end{equation}
The equation we've arrived to is exactly \cref{general-isre-omega} for $\widetilde{\Omega}$, so the solutions must be the same, with $\widetilde{\Omega}=\omega+\frac{i}{\tau}$ for $\omega, \tau \geq 0$, and time dependence $e^{-\frac{t}{\tau}+i\omega t}$. The solutions of the original equation must then have $\Omega = -\widetilde{\Omega}^* = -\omega+\frac{i}{\tau}$ and time dependence $e^{-\frac{t}{\tau}-i\omega t}$; this is merely a reversal of the precession direction.

\section{Validity of the Asymptotic Expression}\label{validity-regimes}
Various approximations were made in the course of the deriving the
asymptotic expressions and first order expressions. The finiteness
and the asymptotic form the of the scattering phase shift has already
been alluded to. Immediately following is the approximation of sine
by its Taylor expansion. Note that the argument being approximated
is not only the phase shift, but it can also be a difference of phase
shifts. This does not essentially change the situation : the factors
of $\pi$ cancel out within the
sine, and the remaining quantities are small for small $k$. 

\begin{figure*}
	\begin{center}
		\includegraphics[width=0.95\textwidth]{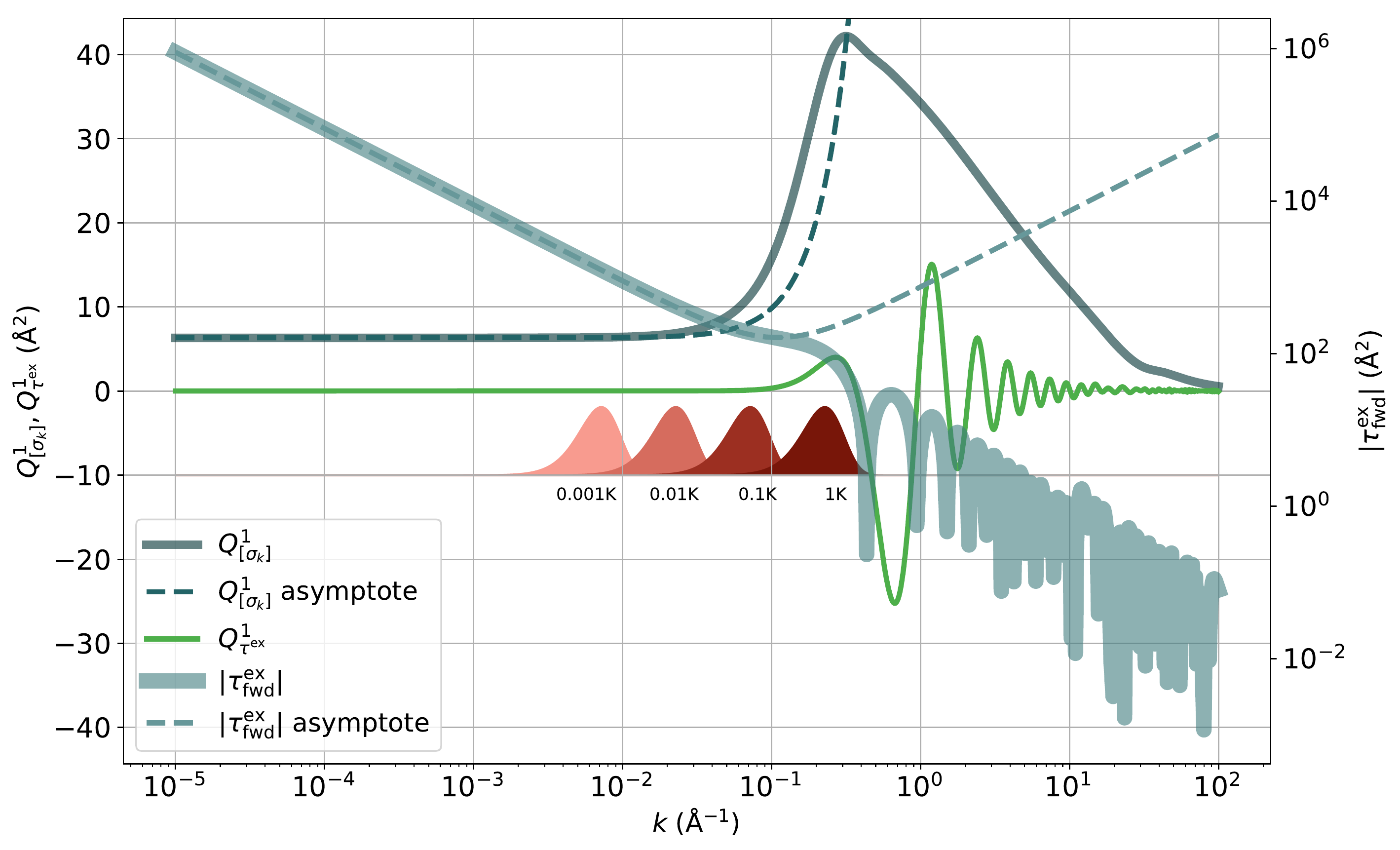}
	\end{center}
	
	\captiondescription{Goodness of Approximations at Different Momenta/Temperatures\label{fig:small-k-approximations}}
        {The figure shows \isreQsigma, \isretauexfwd for atomic hydrogen with their series expansions \cref{eq:q1-sigma-appr,eq:tau-ex-appr}, as well as \isreQtau. These quantities are multiplied with a Gaussian and integrated over all momenta; a few Gaussians for a few different temperatures are shown. For high enough momenta the series expansions naturally break down, so in order for the asymptotic formula to give a good approximation to $\mu$ the Gaussian should be concentrated in the area where the series expansions approximate the quantities well. This gives upper bounds $T_Q$ and $T_{\tau}$ on temperature where the asymptotic formula can be expected to work.}
	
\end{figure*}

The region of validity for these approximations is shown in \cref{fig:small-k-approximations}
for hydrogen. The figure compares \cref{eq:q1-sigma-appr,eq:tau-ex-appr}
with the numerically evaluated expressions for $Q_{[\sigma_{k}]}^{1}$
and $\tau_{\text{fwd}}^{\text{ex}}$; $Q_{[\tau^{\text{ex}}]}^{1}$
is also shown for completeness. The approximations are clearly robust
for $k<\SI{1e-2}{\per\angstrom}$, beyond which they begin to deteriorate.
In the good regime, then, the following conditions should hold
\begin{align}
4\pi k^{2}\left|a_{0}\left(\frac{a_{0}^{3}}{3}+2a_{1}^{3}\right)\right|\ll & 4\pi a_{0}^{2}\Leftrightarrow k^{2}\ll\frac{3\left|a_{0}\right|}{\left|a_{0}^{3}+6a_{1}^{3}\right|}\label{eq:Q1_sigma_small}\\
4\pi k\left|\frac{2}{3}a_{0}^{3}+3a_{1}^{3}\right|\ll & \left|\frac{4\pi a_{0}}{k}\right|\Leftrightarrow k^{2}\ll\frac{3\left|a_{0}\right|}{\left|2a_{0}^{3}+9a_{1}^{3}\right|}\label{eq:tau_ex_fwd_small}
\end{align}

For $\tau_{\text{fwd}}^{\text{ex}}$, $-\frac{4\pi a_{0}}{k}$ seems
to remain a good approximation even beyond the point where the $k$
term begin to deteriorate the approximation (at the turn of the asymptotic
formula). Combined with the overwhelming magnitude of $\tau_{\text{fwd}}^{\text{ex}}$
this may explain why the asymptotic formula remains robust for hydrogen even beyond
the region where these approximations break down. 

In the next step an integration over all momenta is carried out. Obviously
this is in contradiction with the small $k$ approximation needed
for the preceding approximations. However, the functions being integrated
have a Gaussian form, so for suitable parameter regimes the contributions
from higher momenta are negligible. In general the integrals in \cref{eq:omega-alpha}
and \cref{eq:chi-tau} depend on the scaled momentum $\gamma$ as $e^{-\gamma^{2}}\gamma^{q}$;
they are shown for a few temperatures in \cref{fig:small-k-approximations}
($q=5$). For \SI{1}{\kelvin} the approximations are already not
that good, and in fact for hydrogen the approximations seem to be
valid for below \SI{0.01}{\kelvin}. 

Clearly what is needed is for the Gaussian to be centred at a region
where the approximations hold. $e^{-\gamma^{2}}\gamma^{q}$ is centred
at $\gamma^{*}=\sqrt{\frac{q}{2}}=\sqrt{\frac{5}{2}}$, giving $\left(k^{*}\right)^{2}=\frac{5m}{2\beta\hbar^{2}}$,
which should be smaller than the momentum where the approximations
break down. Combining this with \cref{eq:Q1_sigma_small,eq:tau_ex_fwd_small}
results in two conditions for temperature: 
\begin{align}
T\ll & \frac{6\hbar^{2}\left|a_{0}\right|}{5mk_{B}\left|a_{0}^{3}+6a_{1}^{3}\right|} =  T_{Q} \label{eq:T_Q}\\
T\ll & \frac{6\hbar^{2}\left|a_{0}\right|}{5mk_{B}\left|2a_{0}^{3}+9a_{1}^{3}\right|} =  T_{\tau} \label{eq:T_tau}
\end{align}
$T_{\tau}$ is almost always the smaller of the two, except in the narrow region where $-5^{\frac{1}{3}}<\frac{a_0}{a_1}<-3^{\frac{1}{3}}$.

\begin{table}
	\begin{center}
		\begin{tabular}{|c|c|c|c|c|c|}
			\hline 
			Species & $a_{0}$ & $a_{1}$ & $T_{Q}$ & $T_{\tau}$ & $T_{\Lambda}$ \\
			\hline 
			\hline 
			H & \SI{0.71}{\angstrom} & \SI{-2.70}{\angstrom} & \SI{0.34}{\kelvin} & \SI{0.23}{\kelvin} &  \SI{600}{\kelvin}\\
			\hline 
			\ce{^3He} & \SI{-8.1}{\angstrom} & \SI{-3.024}{\angstrom} & \SI{0.22}{\kelvin} & \SI{0.11}{\kelvin} & \SI{1.54}{\kelvin}\\
			\hline 
		\end{tabular}
	\end{center}
	\caption{Small $k$ Regimes\label{tab:validity-temperatures}}
	s-wave and p-wave scattering lengths and the temperature upper bounds $T_Q$ and $T_{\tau}$ (\cref{eq:T_Q,eq:T_tau}) for validity of asymptotic formula for a few atomic species under the triplet potential interaction. $T_{\Lambda}$ is derived form the quantum gas criterion $\frac{\Lambda_{\text{th}}}{\abs{a_s}} > 1$ and is shown for comparison. 
\end{table}

\section{A Dictionary Of Cross Sections}\label{crosssections}
This section largely reproduces definitions and results from \cite{Bouchaud1985-Spinwaves}. However as \cite{Bouchaud1985-Spinwaves} does not give all the required quantities, some of them are derived in \cref{rederivations}.

\Cref{eq:omega-alpha} and \cref{eq:chi-tau}  define $\Omega$ and $\Xi$. In addition another angular average needs to be defined:
\begin{equation}
  \tilde{Q}^t_{\left[ \sigma \right]} = 2 \pi \int_0^{\pi} \sin \theta \sigma \left( \theta \right) \dd{\theta}.
\end{equation}
$\tilde{\Omega}$ uses this quantity instead of $Q$; further,
\begin{equation}
  Q^t_{\left[ \sigma \right]}=\widetilde{Q}^0_{\left[ \sigma \right]}-\widetilde{Q}^t_{\left[ \sigma \right]}.
\end{equation}

What remains is to list the relevant cross sections and angular averages $Q$ in \cref{bouchaud-formulae}. Some of these are given in \cite{Bouchaud1985-Spinwaves}; in addition changes between quantities $\sigma \leftrightarrow \sigma^{\text{ex}}$ can be easily done with the substitution  $(2 L +1) \leftrightarrow (-1)^L (2 L + 1)$. See \cref{rederivations} for the derivation of the rest. The $u$ and $g$ indices in these expressions stand for phase shifts calculated from the triplet and single potentials of hydrogen, respectively.

\begin{widetext}
\begin{align*}
\tilde{Q}_{\left[\sigma_{t}\right]}^{0}= & \frac{\pi}{k^{2}}\sum_{L}(2L+1)\sin^{2}(\delta_{L}^{g}-\delta_{L}^{u})\\
\tilde{Q}_{\left[\sigma_{t}^{ex}\right]}^{0}= & \frac{4\pi}{k^{2}}\sum_{L}\left(-1\right)^{L}(2L+1)\sin^{2}(\delta_{L}^{g}-\delta_{L}^{u})\\
\tilde{Q}_{\left[\sigma_{d}\right]}^{0}= & \frac{\pi}{k^{2}}\sum_{L}\left(2L+1\right)\left[\sin^{2}\delta_{L}^{g}+\sin^{2}\delta_{L}^{u}+2\cos\left(\delta_{L}^{u}-\delta_{L}^{g}\right)\sin\delta_{L}^{u}\sin\delta_{L}^{g}\right]\\
\tilde{Q}_{\left[\sigma_{\mathrm{d}}^{\mathrm{ex}}\right]}^{0}= & \frac{\pi}{k^{2}}\sum_{L}\left(-1\right)^{L}\left(2L+1\right)\left[\sin^{2}\delta_{L}^{g}+\sin^{2}\delta_{L}^{u}+2\cos\left(\delta_{L}^{u}-\delta_{L}^{g}\right)\sin\delta_{L}^{u}\sin\delta_{L}^{g}\right]\\
\tilde{Q}_{[\sigma_{\text{dt}}]}^{0}= & \frac{\pi}{k^{2}}\sum_{L}\left(2L+1\right)\left[\sin^{2}\delta_{L}^{g}-\sin^{2}\delta_{L}^{u}\right]\\
\tilde{Q}_{[\sigma_{\text{dt}}^{\text{ex}}]}^{0}= & \frac{\pi}{k^{2}}\sum_{L}(-1)^{L}\left(2L+1\right)\left[\sin^{2}\delta_{L}^{g}-\sin^{2}\delta_{L}^{u}\right]\\
\tilde{Q}_{[\sigma_{\text{dt}}]}^{1}= & \frac{4\pi}{k^{2}}\sum_{L}(L+1)\left[\sin(\delta_{L+1}^{u})\sin(\delta_{L}^{g})\sin(\delta_{L}^{g}-\delta_{L+1}^{u})+\sin(\delta_{L}^{u})\sin(\delta_{L+1}^{g})\sin(\delta_{L+1}^{g}-\delta_{L}^{u})\right]\\
\tilde{Q}_{[\sigma_{\text{dt}}^{\text{ex}}]}^{1}= & \frac{4\pi}{k^{2}}\sum_{L}(-1)^{L}(L+1)\left[\sin(\delta_{L+1}^{u})\sin(\delta_{L}^{g})\sin(\delta_{L}^{g}-\delta_{L+1}^{u})+\sin(\delta_{L}^{u})\sin(\delta_{L+1}^{g})\sin(\delta_{L+1}^{g}-\delta_{L}^{u})\right]\\
Q_{\left[\sigma_{t}^{ex}\right]}^{0}\equiv & \tilde{Q}_{\left[\sigma_{t}^{ex}\right]}^{0}\\
Q_{\left[\sigma_{d}\right]}^{1}= & \tilde{Q}_{\left[\sigma_{d}\right]}^{0}-\tilde{Q}_{\left[\sigma_{d}\right]}^{1}=\tilde{Q}_{\left[\sigma_{d}\right]}^{0}-0\\
Q_{\left[\sigma_{d}^{ex}\right]}^{1}= & \tilde{Q}_{\left[\sigma_{\mathrm{d}}^{\mathrm{ex}}\right]}^{0}-\tilde{Q}_{\left[\sigma_{\mathrm{d}}^{\mathrm{ex}}\right]}^{1}=\tilde{Q}_{\left[\sigma_{\mathrm{d}}^{\mathrm{ex}}\right]}^{0}-0\\
Q_{\left[\sigma_{dt}\right]}^{1}= & \tilde{Q}_{[\sigma_{\text{dt}}]}^{0}-\tilde{Q}_{[\sigma_{\text{dt}}]}^{1}\\
Q_{\left[\sigma_{dt}^{ex}\right]}^{1}= & \tilde{Q}_{[\sigma_{\text{dt}}^{\text{ex}}]}^{0}-\tilde{Q}_{[\sigma_{\text{dt}}^{\text{ex}}]}^{1}\\
\tilde{Q}_{\left[\tau_{dt}\right]}^{0}= & \frac{2\pi}{k^{2}}\sum_{L}(2L+1)\sin(\delta_{L}^{g})\sin(\delta_{L}^{u})\sin(\delta_{L}^{g}-\delta_{L}^{u})\\
\tilde{Q}_{\left[\tau_{dt}\right]}^{1}= & \frac{2\pi}{k^{2}}\sum_{L}(L+1)\left[\sin(\delta_{L}^{g})\sin(\delta_{L+1}^{g})\sin(\delta_{L+1}^{g}-\delta_{L}^{g})-\sin(\delta_{L}^{u})\sin(\delta_{L+1}^{u})\sin(\delta_{L+1}^{u}-\delta_{L}^{u})\right]\\
\tilde{Q}_{\left[\tau_{t}^{ex}\right]}^{1}= & \frac{2\pi}{k^{2}}\sum_{L}\left(-1\right)^{L}(L+1)\left[\sin(\delta_{L}^{u})\sin(\delta_{L+1}^{u})\sin(\delta_{L+1}^{u}-\delta_{L}^{u})+\sin(\delta_{L}^{g})\sin(\delta_{L+1}^{g})\sin(\delta_{L+1}^{g}-\delta_{L}^{g})\right.\\
- & \left.\sin(\delta_{L}^{g})\sin(\delta_{L+1}^{u})\sin(\delta_{L+1}^{u}-\delta_{L}^{g})-\sin(\delta_{L}^{u})\sin(\delta_{L+1}^{g})\sin(\delta_{L+1}^{g}-\delta_{L}^{u})\right]\\
\tilde{Q}_{\left[\tau_{d}^{ex}\right]}^{1}= & \frac{2\pi}{k^{2}}\sum_{L}\left(-1\right)^{L}(L+1)\left[\sin(\delta_{L}^{u})\sin(\delta_{L+1}^{u})\sin(\delta_{L+1}^{u}-\delta_{L}^{u})+\sin(\delta_{L}^{g})\sin(\delta_{L+1}^{g})\sin(\delta_{L+1}^{g}-\delta_{L}^{g})\right.\\
+ & \left.\sin(\delta_{L}^{g})\sin(\delta_{L+1}^{u})\sin(\delta_{L+1}^{u}-\delta_{L}^{g})+\sin(\delta_{L}^{u})\sin(\delta_{L+1}^{g})\sin(\delta_{L+1}^{g}-\delta_{L}^{u})\right]\\
\tau_{t}^{fwd}= & \frac{\pi}{k^{2}}\sum_{L}(2L+1)\left(\sin(2\delta_{L}^{g})-\sin(2\delta_{L}^{u})\right)\\
\tau_{d}^{bwd}= & \frac{\pi}{k^{2}}\sum_{L}(-1)^{L}(2L+1)\left(\sin(2\delta_{L}^{g})+\sin(2\delta_{L}^{u})\right)\\
\tau_{t}^{bwd}= & \frac{\pi}{k^{2}}\sum_{L}(-1)^{L}(2L+1)\left(\sin(2\delta_{L}^{g})-\sin(2\delta_{L}^{u})\right)
\end{align*}
\end{widetext}
\begin{widetext}
\section{Derivations of Some Quantities}\label{rederivations}
The definitions are once again from \cite{Bouchaud1985-Spinwaves}. The central quantity is the phase-shift expansion of the transition matrix:
\begin{equation}
  T_{\alpha}\left(\vec{k}_{f},\vec{k}_{i}\right)=-\frac{\hbar^{2}}{4\pi^{2}\mu k}\sum_{L}\left(2L+1\right)e^{i\delta_{L}^{\alpha}}\sin\delta_{L}^{\alpha}P_{L}\left(\cos\theta\right).
\end{equation}
Here $\theta$ is the angle between $\vec{k}_{f}$ and $\vec{k}_{i}$, $\alpha$ is either $g$ or $u$, and the $P_L$ are the Legendre polynomials.
One then defines the direct ($d$) and transfer ($t$) matrices as
\begin{align}
  T_{d}&=\frac{1}{2}\left(T_{g}+T_{u}\right) \label{direct-matrix} \\
  T_{t}&=\frac{1}{2}\left(T_{g}-T_{u}\right) \label{transfer-matrix}
\end{align}
and the associated cross sections (with $\alpha, \beta \in \{t,d\}$ and $\sigma_{\alpha \alpha}\equiv \sigma_{\alpha})$
\begin{align}
  \sigma_{\alpha\beta}\left(\theta\right)-i\tau_{\alpha\beta}\left(\theta\right)=&	\frac{16\pi^{4}\mu^{2}}{\hbar^{4}}T_{\alpha}\left(\vec{k}_{f},\vec{k}_{i}\right)T_{\beta}^{*}\left(\vec{k}_{f},\vec{k}_{i}\right) \label{normal-cc} \\
  \sigma_{\alpha\beta}^{\text{ex}}\left(\theta\right)-i\tau_{\alpha\beta}^{\text{ex}}\left(\theta\right)=&\frac{16\pi^{4}\mu^{2}}{\hbar^{4}}T_{\alpha}\left(-\vec{k}_{f},\vec{k}_{i}\right)T_{\beta}^{*}\left(\vec{k}_{f},\vec{k}_{i}\right) \label{exchange-cc}.
\end{align}

\subsection{$\tilde{Q}_{\left[\sigma_{t}^{\text{ex}}\right]}^{1}-i\tilde{Q}_{\left[\tau_{t}^{\text{ex}}\right]}^{1}=0-i\tilde{Q}_{\left[\tau_{t}^{\text{ex}}\right]}^{1}$}
Substituting \cref{transfer-matrix} to \cref{exchange-cc} leads to
\begin{align*}
  \sigma_{t}^{\text{ex}}\left(\theta\right)-i\tau_{t}^{\text{ex}}\left(\theta\right)=&	\frac{16\pi^{4}\mu^{2}}{\hbar^{4}}T_{t}\left(-\vec{k}_{f},\vec{k}_{i}\right)T_{t}^{*}\left(\vec{k}_{f},\vec{k}_{i}\right)\\
=&	\frac{1}{4k^{2}}\sum_{L}\left(2L+1\right)\left[e^{i\delta_{L}^{g}}\sin\delta_{L}^{g}P_{L}\left(\cos\left(\pi-\theta\right)\right)-e^{i\delta_{L}^{u}}\sin\delta_{L}^{u}P_{L}\left(\cos\left(\pi-\theta\right)\right)\right] \\
\times &	\sum_{M}\left(2M+1\right)\left[e^{-i\delta_{M}^{g}}\sin\delta_{M}^{g}P_{M}\left(\cos\theta\right)-e^{-i\delta_{M}^{u}}\sin\delta_{M}^{u}P_{M}\left(\cos\theta\right)\right] \\
= &	\frac{1}{4k^{2}}\sum_{L,M}\left(2L+1\right)\left(2M+1\right)\left(-1\right)^{L}\left[e^{i\left(\delta_{L}^{g}-\delta_{M}^{g}\right)}\sin\delta_{L}^{g}\sin\delta_{M}^{g}P_{L}\left(\cos\theta\right)P_{M}\left(\cos\theta\right)\right. \\
- &e^{i\left(\delta_{L}^{u}-\delta_{M}^{g}\right)}\sin\delta_{L}^{u}\sin\delta_{M}^{g}P_{L}\left(\cos\theta\right)P_{M}\left(\cos\theta\right) \\
- &e^{i\left(\delta_{L}^{g}-\delta_{M}^{u}\right)}\sin\delta_{L}^{g}\sin\delta_{M}^{u}P_{L}\left(\cos\theta\right)P_{M}\left(\cos\theta\right) \\
+ &\left.e^{i\left(\delta_{L}^{u}-\delta_{M}^{u}\right)}\sin\delta_{L}^{u}\sin\delta_{M}^{u}P_{L}\left(\cos\theta\right)P_{M}\left(\cos\theta\right)\right].
\end{align*}
For $\tilde{Q}_{\left[\tau_{t}^{\text{ex}}\right]}^{1}$,  $2\pi\int_{0}^{\pi}\tau_{t}^{\text{ex}}\left(\theta\right)\sin\left(\theta\right)\cos\left(\theta\right)\dd{\theta}$ needs to be calculated. The substitution

\begin{align*}
x=\cos\theta & \Leftrightarrow\arccos\left(x\right)=\theta\\
\dd{x}= & -\sin\theta\dd{\theta}\\
\dd{\theta}= & -\frac{\dd{x}}{\sin\theta}
\end{align*}
leads to the integral

\begin{align*}
 & 2\pi\int_{0}^{\pi}P_{L}\left(\cos\theta\right)P_{M}\left(\cos\theta\right)\sin\left(\theta\right)\cos\left(\theta\right)\dd{\theta}\\
= & 2\pi\int_{1}^{-1}P_{L}\left(x\right)P_{M}\left(x\right)\sin\left(\theta\right)x\left(-\frac{\dd{x}}{\sin\theta}\right)\\
= & 2\pi\int_{-1}^{1}P_{L}\left(x\right)P_{M}\left(x\right)x\dd{x},
\end{align*}
which, using $(L+1)P_{L+1}(x)-(2 L + 1) x P_L(x)+L P_{L-1}(x)=0$ and orthogonality of Legendre polynomials, evaluates to
\begin{align*}
 & 2\pi\int_{-1}^{1}P_{L}\left(x\right)P_{M}\left(x\right)x\dd{x}\\
= & \begin{cases}
2\pi\frac{2\left(L+1\right)}{\left(2L+1\right)\left(2L+3\right)} & \text{for }M=L+1\\
2\pi\frac{2L}{\left(2L-1\right)\left(2L+1\right)} & \text{for }M=L-1.
\end{cases}
\end{align*}

The integrals then give

\begin{align*}
\tilde{Q}_{\left[\sigma_{t}^{\text{ex}}\right]}^{1}-i\tilde{Q}_{\left[\tau_{t}^{\text{ex}}\right]}^{1}= & \frac{4\pi}{4k^{2}}\sum_{L=0}\left(-1\right)^{L}\left(L+1\right)\left[e^{i\left(\delta_{L}^{g}-\delta_{L+1}^{g}\right)}\sin\delta_{L}^{g}\sin\delta_{L+1}^{g}\right.\\
- & e^{i\left(\delta_{L}^{u}-\delta_{L+1}^{g}\right)}\sin\delta_{L}^{u}\sin\delta_{L+1}^{g}-e^{i\left(\delta_{L}^{g}-\delta_{L+1}^{u}\right)}\sin\delta_{L}^{g}\sin\delta_{L+1}^{u}\\
+ & \left.e^{i\left(\delta_{L}^{u}-\delta_{L+1}^{u}\right)}\sin\delta_{L}^{u}\sin\delta_{L+1}^{u}P_{L}\left(\cos\theta\right)P_{L+1}\left(\cos\theta\right)\right]\\
+ & \frac{\pi}{k^{2}}\sum_{L=1}\left(-1\right)^{L}L\left[e^{i\left(\delta_{L}^{g}-\delta_{L-1}^{g}\right)}\sin\delta_{L}^{g}\sin\delta_{L-1}^{g}\right.\\
- & e^{i\left(\delta_{L}^{u}-\delta_{L-1}^{g}\right)}\sin\delta_{L}^{u}\sin\delta_{L-1}^{g}-e^{i\left(\delta_{L}^{g}-\delta_{L-1}^{u}\right)}\sin\delta_{L}^{g}\sin\delta_{L-1}^{u}\\
+ & \left.e^{i\left(\delta_{L}^{u}-\delta_{L-1}^{u}\right)}\sin\delta_{L}^{u}\sin\delta_{L-1}^{u}\right].
\end{align*}

Shifting the latter sum with $M=L-1$ and rearranging leads to
\begin{align}
 & \frac{\pi}{k^{2}}\sum_{L=0}\left(-1\right)^{L}\left(L+1\right)\left[\left(\sin\delta_{L}^{g}\sin\delta_{L+1}^{g}\right)\left(e^{i\left(\delta_{L}^{g}-\delta_{L+1}^{g}\right)}-e^{-i\left(\delta_{L}^{g}-\delta_{L+1}^{g}\right)}\right)\right.\nonumber \\
- & \left(\sin\delta_{L}^{u}\sin\delta_{L+1}^{g}\right)\left(e^{i\left(\delta_{L}^{u}-\delta_{L+1}^{g}\right)}-e^{-i\left(\delta_{L}^{u}-\delta_{L+1}^{g}\right)}\right)\nonumber \\
- & \ldots\nonumber \\
= & \frac{2\pi}{k^{2}}i\sum_{L=0}\left(-1\right)^{L}\left(L+1\right)\left[\left(\sin\delta_{L}^{g}\sin\delta_{L+1}^{g}\right)\sin\left(\delta_{L}^{g}-\delta_{L+1}^{g}\right)\right.\nonumber \\
- & \left(\sin\delta_{L}^{u}\sin\delta_{L+1}^{g}\right)\sin\left(\delta_{L}^{u}-\delta_{L+1}^{g}\right)\nonumber \\
- & \left(\sin\delta_{L}^{g}\sin\delta_{L+1}^{u}\right)\sin\left(\delta_{L}^{g}-\delta_{L+1}^{u}\right)\nonumber \\
  + & \left.\left(\sin\delta_{L}^{u}\sin\delta_{L+1}^{u}\right)\sin\left(\delta_{L}^{u}-\delta_{L+1}^{u}\right)\right]\label{qt1_sigma_tau_ex}\\
= & 0-i\tilde{Q}_{\left[\tau_{t}^{\text{ex}}\right]}^{1}\nonumber, 
\end{align}

and finally 
\begin{align*}
\tilde{Q}_{\left[\tau_{t}^{\text{ex}}\right]}^{1}= & \frac{2\pi}{k^{2}}\sum_{L=0}\left(-1\right)^{L}\left(L+1\right)\left[\left(\sin\delta_{L}^{g}\sin\delta_{L+1}^{g}\right)\sin\left(\delta_{L+1}^{g}-\delta_{L}^{g}\right)\right.\\
- & \left(\sin\delta_{L}^{u}\sin\delta_{L+1}^{g}\right)\sin\left(\delta_{L+1}^{g}-\delta_{L}^{u}\right)\\
- & \left(\sin\delta_{L}^{g}\sin\delta_{L+1}^{u}\right)\sin\left(\delta_{L+1}^{u}-\delta_{L}^{g}\right)\\
+ & \left.\left(\sin\delta_{L}^{u}\sin\delta_{L+1}^{u}\right)\sin\left(\delta_{L+1}^{u}-\delta_{L}^{u}\right)\right] .
\end{align*}

\subsection{$\tilde{Q}_{\left[\sigma_{d}\right]}^{1}-i\tilde{Q}_{\left[\tau_{d}\right]}^{1}=0-i\tilde{Q}_{\left[\tau_{d}\right]}^{1}$} \label{Qt1_d}

This is just \cref{qt1_sigma_tau_ex} with two modifications. Changing
transfer to direct ($t\to d$) only affects the signs of the middle
terms, flipping them from minus to plus. $\sigma^{\text{ex}}\to\sigma$
means leaving out the $(-1)^{L}$. Hence the final expression is $\tilde{Q}_{\left[\sigma_{d}\right]}^{1}=0$ and

\begin{align*}
\tilde{Q}_{\left[\tau_{d}\right]}^{1}= & \frac{2\pi}{k^{2}}\sum_{L=0}\left(L+1\right)\left[\left(\sin\delta_{L}^{g}\sin\delta_{L+1}^{g}\right)\sin\left(\delta_{L+1}^{g}-\delta_{L}^{g}\right)\right.\\
+ & \left(\sin\delta_{L}^{u}\sin\delta_{L+1}^{g}\right)\sin\left(\delta_{L+1}^{g}-\delta_{L}^{u}\right)\\
+ & \left(\sin\delta_{L}^{g}\sin\delta_{L+1}^{u}\right)\sin\left(\delta_{L+1}^{u}-\delta_{L}^{g}\right)\\
+ & \left.\left(\sin\delta_{L}^{u}\sin\delta_{L+1}^{u}\right)\sin\left(\delta_{L+1}^{u}-\delta_{L}^{u}\right)\right] .
\end{align*}

\subsection{$\tilde{Q}_{\left[\sigma_{d}\right]}^{0}+i\tilde{Q}_{\left[\tau_{d}\right]}^{0}=\tilde{Q}_{\left[\sigma_{d}\right]}^{0}+i0$}
The scattering cross section expression comes about the same way as in \cref{Qt1_d}.

\begin{align*}
\sigma_{d}\left(\theta\right)-i\tau_{d}\left(\theta\right)= & \frac{16\pi^{4}\mu^{2}}{\hbar^{4}}T_{d}\left(\vec{k}_{f},\vec{k}_{i}\right)T_{d}^{*}\left(\vec{k}_{f},\vec{k}_{i}\right)\\
= & \frac{1}{4k^{2}}\sum_{L}\left(2L+1\right)\left[e^{i\delta_{L}^{g}}\sin\delta_{L}^{g}P_{L}\left(\cos\theta\right)+e^{i\delta_{L}^{u}}\sin\delta_{L}^{u}P_{L}\left(\cos\theta\right)\right]\\
\times & \sum_{M}\left(2M+1\right)\left[e^{-i\delta_{M}^{g}}\sin\delta_{M}^{g}P_{M}\left(\cos\theta\right)+e^{-i\delta_{M}^{u}}\sin\delta_{M}^{u}P_{M}\left(\cos\theta\right)\right]\\
= & \frac{1}{4k^{2}}\sum_{L,M}\left(2L+1\right)\left(2M+1\right)\left[e^{i\left(\delta_{L}^{g}-\delta_{M}^{g}\right)}\sin\delta_{L}^{g}\sin\delta_{M}^{g}P_{L}\left(\cos\theta\right)P_{M}\left(\cos\theta\right)\right.\\
+ & e^{i\left(\delta_{L}^{u}-\delta_{M}^{g}\right)}\sin\delta_{L}^{u}\sin\delta_{M}^{g}P_{L}\left(\cos\theta\right)P_{M}\left(\cos\theta\right)\\
+ & e^{i\left(\delta_{L}^{g}-\delta_{M}^{u}\right)}\sin\delta_{L}^{g}\sin\delta_{M}^{u}P_{L}\left(\cos\theta\right)P_{M}\left(\cos\theta\right)\\
+ & \left.e^{i\left(\delta_{L}^{u}-\delta_{M}^{u}\right)}\sin\delta_{L}^{u}\sin\delta_{M}^{u}P_{L}\left(\cos\theta\right)P_{M}\left(\cos\theta\right)\right]
\end{align*}  

For this quantity the integral is simpler, in fact is it just the orthogonality relation for Legendre polynomials:
\begin{align*}
 & 2\pi\int_{0}^{\pi}P_{L}\left(\cos\theta\right)P_{M}\left(\cos\theta\right)\sin\left(\theta\right)\dd{\theta}\\
= & 2\pi\int_{1}^{-1}P_{L}\left(x\right)P_{M}\left(x\right)\sin\left(\theta\right)\left(-\frac{\dd{x}}{\sin\theta}\right)\\
= & 2\pi\int_{-1}^{1}P_{L}\left(x\right)P_{M}\left(x\right)\dd{x}\\
= & 2\pi\frac{2}{2L+1}\delta_{M}^{L}.
\end{align*}

Hence the result: 
\begin{align}
 & =\frac{\pi}{k^{2}}\sum_{L}\left(2L+1\right)\left[\sin\delta_{L}^{g}\sin\delta_{L}^{g}\right.\nonumber \\
+ & e^{i\left(\delta_{L}^{u}-\delta_{L}^{g}\right)}\sin\delta_{L}^{u}\sin\delta_{L}^{g}+e^{i\left(\delta_{L}^{g}-\delta_{L}^{u}\right)}\sin\delta_{L}^{g}\sin\delta_{L}^{u}\nonumber \\
+ & \left.\sin\delta_{L}^{u}\sin\delta_{L}^{u}\right]\label{qt0_sigma_tau_d}\\
= & \frac{\pi}{k^{2}}\sum_{L}\left(2L+1\right)\left[\sin^{2}\delta_{L}^{g}+\sin^{2}\delta_{L}^{u}+2\cos\left(\delta_{L}^{u}-\delta_{L}^{g}\right)\sin\delta_{L}^{u}\sin\delta_{L}^{g}\right] \label{qt0_sigma_tau_d-result} \\
= & \tilde{Q}_{\left[\sigma_{d}\right]}^{0}+i0\nonumber . 
\end{align}

\subsection{$Q_{\left[\sigma_{d}\right]}^{1}=\tilde{Q}_{\left[\sigma_{d}\right]}^{0}-\tilde{Q}_{\left[\sigma_{d}\right]}^{1}=\tilde{Q}_{\left[\sigma_{d}\right]}^{0}$}
In \cref{Qt1_d}, it was shown that $\tilde{Q}_{\left[\sigma_{d}\right]}^{1}=0$. The result is then given by \cref{qt0_sigma_tau_d-result}.

\subsection{$\tilde{Q}_{\left[\sigma_{\text{dt}}\right]}^{0}-i\tilde{Q}_{\left[\tau_{\text{dt}}\right]}^{0}$} \label{Qt0_dt}
The cross sections are given by

\begin{align*}
\sigma_{\text{dt}}\left(\theta\right)-i\tau_{\text{dt}}\left(\theta\right)= & \frac{16\pi^{4}\mu^{2}}{\hbar^{4}}T_{d}\left(\vec{k}_{f},\vec{k}_{i}\right)T_{t}^{*}\left(\vec{k}_{f},\vec{k}_{i}\right)\\
= & \frac{1}{4k^{2}}\sum_{L}\left(2L+1\right)\left[e^{i\delta_{L}^{g}}\sin\delta_{L}^{g}P_{L}\left(\cos\left(\pi-\theta\right)\right)+e^{i\delta_{L}^{u}}\sin\delta_{L}^{u}P_{L}\left(\cos\left(\pi-\theta\right)\right)\right]\\
\times & \sum_{M}\left(2M+1\right)\left[e^{-i\delta_{M}^{g}}\sin\delta_{M}^{g}P_{M}\left(\cos\theta\right)-e^{-i\delta_{M}^{u}}\sin\delta_{M}^{u}P_{M}\left(\cos\theta\right)\right]\\
= & \frac{1}{4k^{2}}\sum_{L,M}\left(2L+1\right)\left(2M+1\right)\left[e^{i\left(\delta_{L}^{g}-\delta_{M}^{g}\right)}\sin\delta_{L}^{g}\sin\delta_{M}^{g}P_{L}\left(\cos\theta\right)P_{M}\left(\cos\theta\right)\right.\\
+ & e^{i\left(\delta_{L}^{u}-\delta_{M}^{g}\right)}\sin\delta_{L}^{u}\sin\delta_{M}^{g}P_{L}\left(\cos\theta\right)P_{M}\left(\cos\theta\right)\\
- & e^{i\left(\delta_{L}^{g}-\delta_{M}^{u}\right)}\sin\delta_{L}^{g}\sin\delta_{M}^{u}P_{L}\left(\cos\theta\right)P_{M}\left(\cos\theta\right)\\
- & \left.e^{i\left(\delta_{L}^{u}-\delta_{M}^{u}\right)}\sin\delta_{L}^{u}\sin\delta_{M}^{u}P_{L}\left(\cos\theta\right)P_{M}\left(\cos\theta\right)\right].
\end{align*}

This is just \cref{qt0_sigma_tau_d} with a few minus signs switching
place. Hence

\begin{align*}
= & \frac{\pi}{k^{2}}\sum_{L}\left(2L+1\right)\left[\sin\delta_{L}^{g}\sin\delta_{L}^{g}\right.\\
+ & e^{i\left(\delta_{L}^{u}-\delta_{L}^{g}\right)}\sin\delta_{L}^{u}\sin\delta_{L}^{g}\\
- & e^{i\left(\delta_{L}^{g}-\delta_{L}^{u}\right)}\sin\delta_{L}^{g}\sin\delta_{L}^{u}\\
- & \left.\sin\delta_{L}^{u}\sin\delta_{L}^{u}\right]\\
 & =\frac{\pi}{k^{2}}\sum_{L}\left(2L+1\right)\left[\sin^{2}\delta_{L}^{g}-\sin^{2}\delta_{L}^{u}+2i\sin\left(\delta_{L}^{u}-\delta_{L}^{g}\right)\sin\delta_{L}^{u}\sin\delta_{L}^{g}\right]\\
 & =\tilde{Q}_{\left[\sigma_{dt}\right]}^{0}-i\tilde{Q}_{\left[\tau_{dt}\right]}^{0},
\end{align*}

and finally 
\begin{align*}
\tilde{Q}_{\left[\sigma_{dt}\right]}^{0}= & \frac{\pi}{k^{2}}\sum_{L}\left(2L+1\right)\left[\sin^{2}\delta_{L}^{g}-\sin^{2}\delta_{L}^{u}\right]\\
\tilde{Q}_{\left[\tau_{dt}\right]}^{0}= & \frac{\pi}{k^{2}}\sum_{L}2\left(2L+1\right)\sin\left(\delta_{L}^{u}-\delta_{L}^{g}\right)\sin\delta_{L}^{u}\sin\delta_{L}^{g}
\end{align*}

\subsection{$Q_{\left[\sigma_{\text{dt}}\right]}^{1}==\tilde{Q}_{\left[\sigma_{\text{dt}}\right]}^{0}-\tilde{Q}_{\left[\sigma_{\text{dt}}\right]}^{1}$}

\cite{Bouchaud1985-Spinwaves} gives $\tilde{Q}_{\left[\sigma_{\text{dt}}\right]}^{1}$, and $\tilde{Q}_{\left[\sigma_{\text{dt}}\right]}^{0}$ was calculated in \cref{Qt0_dt}.

\end{widetext}

\bibliographystyle{iopart-num}
\bibliography{asymptotic-mu}

\end{document}